\documentclass[12pt]{iopart}

\usepackage{graphicx}

\usepackage{array}

\def\be{\begin{equation}}
\def\ee{\end{equation}}
\def\bea{\begin{eqnarray}}
\def\eea{\end{eqnarray}}
\def\ba{\begin{array}}
\def\ea{\end{array}}
\def\bdm{\begin{displaymath}}
\def\edm{\end{displaymath}}

\begin{document}

\title[Pseudospin bases for Cu:Bi$_2$Se$_3$ ]
{Pseudospin bases for a model of Cu:Bi$_2$Se$_3$}

\author{Sungkit Yip}

\address{Institute of Physics and Institute of Atomic and Molecular Sciences\\
 Academia Sinica \\
128 Academia Road, Sec 2, Nankang,
Taipei 115, Taiwan.}
\ead{yip@phys.sinica.edu.tw}
\begin{abstract}

We consider the construction of pseudospin bases for
a time-reversal and inversion symmetric system, illustrated
by a model for Cu:Bi$_2$Se$_3$.  Different methods and
bases are compared.

\noindent 

\end{abstract}

\maketitle


\section{Introduction}\label{sec:intro}

Bi$_2$Se$_3$, a well-established topological insulator, becomes
superconducting when doped or intercalated with copper
\cite{Hor10,Wray10,Kriener11} or other elements such as strontium \cite{Sr}.
There is much debate on the superconducting state for this material.
Initially the favorite was the
fully gapped odd-parity state with A$_{1u}$ symmetry
respecting all rotational symmetries of the crystal
\cite{Kriener11,FuBerg10,Sasaki11}.
However,
possibilities later considered include also conventional s-wave
\cite{Levy12,Peng13},
and more recently, broken rotational symmetry odd parity
states \cite{Fu14,VKF16,NMR,Hc2,Yonezawa}.
Moreover,  the superconducting properties of this material
have some unusual characteristics
\cite{Kriener12,Sandilands14}.
The mechanism for superconductivity is also hotly debated
\cite{FuBerg10,Brydon14,Wan14}.

Early on, theoretical works on superconductivity in this material
\cite{FuBerg10,Hao11,Hsieh12,Yamakage12}
started with
effective multi-orbital model motivated by the topological insulator
Bi$_2$Se$_3$ itself.   Thus the minimum model of the normal state
part of the Hamiltonian $H_N$ contains
four degrees of freedom, resulting from two orbitals of different
parity together with two spins degree of freedom.
For example, we can write, following \cite{FuBerg10}
  \be
  H_N (\vec k) = m \sigma_x + v_z k_z \sigma_y + v \sigma_z (k_x s_y - k_y s_x)
  \label{HN}
  \ee
 where $\sigma_z = \pm 1$ represents the two
  (mainly) $p_z$ orbitals in the quintuple layer of Bi$_2$Se$_3$,
  and $\vec s$ the spin.
    Here $\vec k$, $k_x, k_y, k_z$ represent the wavevector and its components,
  $v_z$, $v$ are velocities. $m$ is a quantity of dimension energy
  distinguishing the topologically trivial (commonly taken as $m>0$)
  and the non-trivial cases ($m<0$).  For simplicity, we shall
  confine ourselves to the vicinity of $\vec k = 0$.
(This model is thus valid only for very small doping.  The main point we would
like to discuss in this paper is the methodology of finding
the pseudospin basis. Complications such as higher order
terms in $\vec k$, as well as the fact that the Fermi surface seemingly
changes shape substantially for larger doping \cite{Lahoud13}, would
not be considered here.
 As noted before (e.g. \cite{Yip13}), the model \eref{HN} thus
has symmetry $D_{\infty, h}$ which is actually higher than $D_{3d}$ of
the crystal. )
An important feature of this model is that spins and orbital
degrees of freedom are coupled, as this is necessary to describe
the topological insulator.
Expressing superconducting pairing also in terms of these orbitals and spins,
properties of the superconductor can be directly evaluated
(e.g. \cite{FuBerg10,Hao11,Hsieh12,Yamakage12}).
While this approach is perfectly alright by itself, the
connection with the previous superconductivity literature
\cite{Anderson84,VG84,Ueda85,Blount85,Sigrist91,Yip93,Joynt02,Yip14},
in particular those on heavy fermion superconductivity where
spin-orbit coupling also cannot be ignored, is unclear.
There the approach was to express electronic states near
the Fermi surface in the normal state in terms of pseudospins,
(note that these then depend only on the normal state Hamiltonian),
and Cooper-pairing between
pseudospin states at opposite momenta then considered.
One can connect these two pictures by, starting from say
the multi-orbital model, explicitly construct this
pseudospin basis set for the electrons near the Fermi surface,
and then express the pairing in terms of that between the pseudospins
at $\pm \vec k$.
For the basis set to be ``more useful" in the sense that
connection with the earlier literature
\cite{Anderson84,VG84,Ueda85,Blount85,Sigrist91,Yip93,Joynt02,Yip14}
can be made, these pseudospin states must satisfy
certain symmetry criteria, in particular those imposed
by time-reversal, parity and rotational symmetries.

At a typical point (labeled by the momentum $\vec k$)
on the Fermi surface,
there are two degenerate states since the crystal obeys time-reversal
and parity symmetries.  Given the Hamiltonian,
it is straight-forward to find
two orthonormal states, say
$|\vec k, \alpha' \rangle$ and $|\vec k, \beta' \rangle$,
 at each point on the Fermi surface.
Symmetry operations map states from one point of the Fermi surface
to a linear combination of those at another (and for some
operations, the same) point.
The task at hand is to find
a suitable unitary transformation
among these two states $|\vec k, \alpha' \rangle$ and $|\vec k, \beta' \rangle$
to form two new ones $|\vec k, \alpha \rangle$ and $|\vec k, \beta \rangle$
so that the latter would transform the same
way as ``up" and ``down" spins under the symmetry operations.
\footnote{Actually, to discuss superconductivity,
it is sufficient that the Cooper pairs have
the proper transformations:  see, e.g., \cite{Blount85}.  Hence
our construction here has imposed more constraints than is necessary.
However, we shall keep to this more stringent requirement for
easier presentation.}\label{s-vs-p}
One such explicit construction for the model \eref{HN}
was given by the author himself \cite{Yip13},
and another, slightly later, by Fu and his coworkers \cite{Fu15,KF15,VKF16}.
The two pseudospin bases differ slightly, reflecting the
non-uniqueness of their choice.

With these constructions, one can, for example, now re-expressed
the pairings $\Delta_{1,2,3,4}$
introduced in \cite{FuBerg10}, originally in terms of orbitals and spins
of \eref{HN},
in terms of the basis functions listed in, e.g.
\cite{Blount85,Sigrist91,Yip93}.
\footnote{Rigorously speaking, the symmetry of Bi$_2$Se$_3$ is
$D_{3d}$, and Cooper-pair basis functions for this symmetry were not listed
in the above references.  However, since $D_{3d}$ is a subgroup of
$D_{6h}$, the pair basis functions can be read off from these references
by identifying $B_1$ with $A_1$, $B_2$ with $A_2$, and
$E_2$ with $E_1$ (separately for even and odd parities $g$ and $u$).
For example, in $E_u$, basis functions listed under
$E_{1u}$ and $E_{2u}$ are both acceptable
(though some care is needed to identify the correct partners,
c.f., e.g., \cite{Yip14}).}
Many properties of these superconducting states are then directly evident.
Some examples are given in \cite{Yip13,VKF16} (see also \cite{others}).

The purpose of the present paper is to discuss the relation
between these two methods (\cite{Yip13} on one hand and
\cite{Fu15,KF15,VKF16} on the other)
of constructing the pseudospins.
In the beginning of the next section,
%
we first review the essence of both methods.  We then illustrate
in \sref{subsec:m1} that a slight modifications of \cite{Yip13}
would amount to a method and hence results identical to \cite{Fu15,KF15,VKF16}.
We then construct in \sref{subsec:m2} yet another basis
based on the same principle in \cite{Yip13}.
The advantages and disadvantages of the different
pseudospin bases are discussed.

\section{Review of the methods in \cite{Yip13} and
\cite{Fu15,KF15,VKF16}}\label{sec:2}

We briefly review the methods in  \cite{Yip13} and
\cite{Fu15,KF15,VKF16}.  We shall do this
in the context of the two-orbital two-spin model of \eref{HN},
though the basic points that we are making are completely general.
For (\ref{HN}), the eigenstates
$|\vec k, \alpha' \rangle$ and $|\vec k, \beta' \rangle$
can be represented as column vectors of $4$ entries.
The transformation relations, under
rotations say, among these states, can look unfamiliar
and the construction of
$|\vec k, \alpha \rangle$ and $|\vec k, \beta \rangle$ not immediately obvious.
To overcome this point, the author \cite{Yip13}
made use of the fact that the magnetic moment must transform
like a pseudo-vector.  Pretending that this magnetic moment simply
arises from the spin, he was led to
consider the spin operator $\vec s$ within the two-dimensional
space generated by $|\vec k, \alpha' \rangle$ and $|\vec k, \beta' \rangle$.
Thus $s_{x,y,z}$ become $2 \times 2$ matrices, and are
 then linear combinations of
 $\rho'_{x,y,z}$, the Pauli matrices in
 $|\vec k, \alpha' \rangle$ and $|\vec k, \beta' \rangle$ space.
  In general  $\rho'_{x,y,z}$ would fail to transform as
  the $x,y,z$ components of a pseudo-vector.
  The required proper basis
 $|\vec k, \alpha \rangle$ and $|\vec k, \beta \rangle$,
can be easily identified, since
the Pauli matrices $\rho_{x,y,z}$ in this space is related to
$\rho'_{x,y,z}$ by a rotation which is related to the unitary
transformation between
 $|\vec k, \alpha \rangle$, $|\vec k, \beta \rangle$
 and  $|\vec k, \alpha' \rangle$, $|\vec k, \beta' \rangle$.
As a byproduct, one also obtains an expression of the magnetic
moment of quasiparticles in terms of the pseudospins. \cite{Yip13}

Actually, the general form for the magnetic moment corresponding to this
two-orbital model is more complicated.  As pointed out in
\cite{Liu10}, the magnetic moment, in the notations of
\eref{HN}, is a linear combination of $\vec s$ and
$\sigma_x \vec s$ (more precise statements below).
One is actually free to choose any linear combination between these
two quantities to carry out the procedure outlined above,
since they are both pseudovectors.
A slightly different pseudospin basis would be obtained,
and this reflects the non-uniqueness of the choice of
the pseudospin basis.

On the other hand, Fu and his coworkers \cite{Fu15,KF15,VKF16}
introduce what they
called ``manifestly covariant Bloch basis'' (MCBB).  They
construct this by first identifying a point which they
regard as invariant under the relevant symmetries.  By
demanding that
$|\vec k, \alpha \rangle$ and $|\vec k, \beta \rangle$
at this point be proportional to a real number times
a wavefunction consisting of spin up and down respectively,
they show that this basis automatically possesses the
required transformation, in particular rotational properties.
For the example of \eref{HN}, they regard the system as
invariant under the parity operation $\sigma_x$.  In this way,
a pseudospin basis was constructed \cite{Fu15,KF15,VKF16}.

A moment of thought shows that, in the method used by
the author \cite{Yip13}, if the operator $( 1 + \sigma_x) \vec s$
was employed instead of $\vec s$ in \cite{Yip13}, the basis
constructed would then just be MCBB. (Lest this is not
obvious, an explicit demonstration would be given in Sec \ref{subsec:m1}.).
Viewed entirely mathematically, in both \cite{Yip13} and \cite{Fu15,KF15,VKF16},
the procedures
used amount to reducing the effective wavefunctions to be
only column matrices of two entries.  In \cite{Yip13}, it
was done by choosing a pseudovector operator and evaluating it
in the Hilbert space $|\vec k, \alpha' \rangle$ and $|\vec k, \beta' \rangle$,
and one worked with $2 \times 2$ matrices rather than wavefunctions.
In \cite{Fu15,KF15,VKF16},
it was done by identifying a particular symmetry operation that does so.
The choice needed for the procedure in \cite{Yip13} can always
be chosen as $\vec s$ or any suitable modifications.
The same choice can be used for more general situations,
such as when the model
contains more orbitals than two, or when the symmetry operations
contain screw axes or glide planes.   For \cite{Fu15,KF15,VKF16},
the parity operator may not directly be able to reduce the
wavefunctions to a two-component one, though one is likely
able to identify a suitable generalization.

\subsection{MCBB via the method of \cite{Yip13}}\label{subsec:m1}

Now we turn to explicit calculations to verify our claim in
the last paragraph, that is, the pseudospins constructed using
the method of \cite{Yip13} with a slightly different operator
would yield directly the MCBB pseudospins of \cite{Fu15,KF15,VKF16}.
As mentioned, for \eref{HN},  generally the magnetic moment $\vec m$ is a linear
combination $\vec s$ and $\vec \sigma_x \vec s$.
More precisely, its components can be written as,
\bea
m_z &=&
g_{1z} \frac{1+\sigma_x}{2} s_z +  g_{2z} \frac{1-\sigma_x}{2} s_z
\label{mz} \\
m_{x,y} &=&
g_{1p} \frac{1+\sigma_x}{2} s_{x,y} +  g_{2p} \frac{1-\sigma_x}{2} s_{x,y}
\label{mxy}
\eea
These equations are simply eq (45) of \cite{Liu10} adopted to the
notations here of \eref{HN}. (Our $g_{1z} ... g_{2p}$ are thus the same as
those in \cite{Liu10}).
 For convenience,
we have dropped factors of $\frac{\mu_B}{2}$ where $\mu_B$ is the Bohr magneton
to simplify \eref{mz}, \eref{mxy} and other equations for $\vec m$ below.
We shall write $m_z = m_{1z} + m_{2z}$  where $m_{1z}$ and $m_{2z}$
are the two contributions proportional to $g_{1z}$ and $g_{2z}$
in \eref{mz} and similarly
for $m_x$ and $m_y$ for \eref{mxy}.

We now construct the pseudospin basis for the model \eref{HN}
using the same procedure as in \cite{Yip13} with only the difference
that $\vec m_{1}$ would be used rather than $\vec s$ there.
Let us first recall that it is sufficient to construct the pseudospins
for half of the Fermi sphere (which we shall refer to
as ``northern hemisphere"), since we always require \cite{Yip13},
for any given $\vec k$,
\be
| - \vec k, \alpha \rangle = P | \vec k, \alpha \rangle
\label{P}
\ee
and
\be
|\vec k, \beta \rangle = TP | \vec k, \alpha \rangle
\label{TP}
\ee
(hence also $| - \vec k, \beta \rangle = P | \vec k, \beta \rangle$)
where $P,T$ are the parity and time-reversal operators respectively.
We shall adopt
    the sign convention $ T | s_z=1 \rangle = | s_z = -1 \rangle $
    (thus $ T | s_z= -1 \rangle = - | s_z = 1 \rangle $) for the
    spin wavefunctions.
For the evaluations below, it is useful to note that
$P_+ \equiv \frac{1+\sigma_x}{2}$ is a projection operator:
$P_+^2 = P_+$, and we have $m_{1z} = g_{1z} P_+ s_z$ and similarly for
the $x$ and $y$ components.
For a given point $\vec k$ on the Fermi surface, let us start with
the two degenerate solutions \cite{Yip13} at $\vec k$:
    \be
|\vec k, \alpha'> \equiv \frac{1}{\sqrt{2} \mathcal{N}_{\vec k}}
    e^{ i \vec k \cdot \vec r}
    \left( \ba{c} E_{\vec k} + v k_\| \\
     m + i v_z k_z \ea \right)
    \left( \ba{c}
   1 \\
   i e^{i \phi_{\vec k}} \ea \right)
   \label{ka1}
   \ee
       \be
    |\vec k, \beta'> \equiv \frac{1}{\sqrt{2} \mathcal{N}_{\vec k}}
    e^{ i \vec k \cdot \vec r}
    \left( \ba{c} m - i v_z k_z  \\
      E_{\vec k} + v k_\| \ea \right)
    \left( \ba{c}
   i e^{ - i \phi_{\vec k}} \\
    1 \ea \right)
    \label{kb1}
    \ee
    We are using the notation that the first column matrix denotes
    the part in orbital space
   and the second part denotes the spin space. Here $E_{\vec k}$
   is the energy of the particle
   (which is $( m^2 + v_z^2 k_z^2 + v^2 k_\|^2)^{1/2}$ for
   positive energies), and
    $\mathcal{N}_{\vec k} \equiv [ 2 E_{\vec k} ( E_{\vec k} + v k_\|)]^{1/2}$
   is a renormalization factor.  Here $k_\|$, $\phi_{\vec k}$ are
   respectively the magnitude and azimuthal angle of
   the momentum in the x-y plane.
   Note that we have already chosen $|\vec k, \beta' \rangle = TP
    |\vec k, \alpha' \rangle$.

    Following the procedure in \cite{Yip13}, we shall evaluate
    $\vec m_1$ and hence $ P_1 \vec s$ in the
    $| \vec k, \alpha' \rangle$ $| \vec k, \beta' \rangle$ space.
    It is convenient to first evaluate
    $P_+ | \vec k, \alpha' \rangle$ and $P_+ | \vec k, \beta' \rangle$,
    since for example
    $\langle \vec k, \alpha' | P_+ \vec s | \vec k, \alpha' \rangle$
     = $\langle \vec k, \alpha' | P_+ \vec s P_+ | \vec k, \alpha' \rangle$.
     We easily find (suppressing the plane wavefactors
        $e^{ i \vec k \cdot \vec r}$ from now on)
        \be
    P_+ |\vec k, \alpha'> = \frac{W_{\vec k}}{\sqrt{2}}
    \left( \ba{c} 1 \\
     1 \ea \right)
    \left( \ba{c}
   1 \\
   i e^{i \phi_{\vec k}} \ea \right)
   \label{P1ka1}
   \ee
   and
         \be
    P_+ |\vec k, \beta'> = \frac{W_{\vec k}}{\sqrt{2}}
    \left( \ba{c} 1 \\
     1 \ea \right)
    \left( \ba{c}
   i e^{-i \phi_{\vec k}} \\
    1 \ea \right)
    \label{P1kb1}
    \ee
   Here
   \be
   W_{\vec k} \equiv \frac{(E_{\vec k} + v k_\|) +
     (m + i v_z k_z)}{\sqrt{2} \mathcal{N}_{\vec k}}
     \label{W}
     \ee
     is a coefficient resulting from the projection $P_+$.
     We easily get
     \be
     |W_{\vec k}| = [\frac{1}{2} ( 1 + \frac{m}{E_{\vec k}})] ^{1/2} \ .
     \label{Wmag}
     \ee
     If we consider the branch of \eref{HN} with $E_{\vec k} > 0$, then
     for $m > 0$, $|W_{\vec k = 0}|=1$, which reflects that this state
     corresponds to even parity in the model of \eref{HN}.
     If $m < 0$, then $|W_{\vec k = 0}| = 0$.
     At the chemical potential $\mu$,
     $|W_{\vec k}| = [\frac{1}{2} ( 1 + \frac{m}{\mu})] ^{1/2}$.
     (For more general models than eq (\ref{HN}), $|W_{\vec k}|$
     can still depend on the direction of $\vec k$ even on the Fermi surface.)
     For Bi$_2$Se$_3$ which has band inversion $m < 0$,
     $|W| < 1/\sqrt{2} $ at the Fermi level if it is electron-doped ($\mu > 0$).
     To simplify our notations here and below,
     we shall leave out the subscript $\vec k$ for $W_{\vec k}$
     whenever no confusion can arise.

   Note that
     \be
   | \hat k + > = \frac{1}{\sqrt{2}}
   \left( \ba{c}
   1 \\
   i e^{i \phi_{\vec k}} \ea \right)
   \label{k+}
   \ee
   \be
   |\hat k - > = \frac{1}{\sqrt{2}}
   \left( \ba{c}
   i e^{-i \phi_{\vec k}} \\ 1 \ea \right)
   \label{k-}
   \ee
   are respectively the spin wavefunctions for spins along
   $\pm \hat z \times \hat k$.
   Since $\langle \hat k \pm | s_z | \hat k \pm \rangle = 0$
   whereas $\langle \hat k \mp | s_z | \hat k \pm \rangle =
   \mp i  e^{ \pm i \phi_{\vec k}}$, we easily find that
   $m_{1z}$ has the following matrix form in
     $| \vec k, \alpha' \rangle$ $| \vec k, \beta' \rangle$ space:
        \bea
   m_{1z} (\vec k) &\to& g_{1z}
   \left( \ba{cc} < \vec k, \alpha'| P_1 s_{1z} | \vec k, \alpha'> &
      < \vec k, \alpha'|P_1 s_{1z} | \vec k, \beta'> \\
      < \vec k, \beta'| P_1 s_{1z} | \vec k, \alpha'> &
      < \vec k, \beta'| P_1 s_{1z} | \vec k, \beta'> \ea
      \right)  \nonumber \\
      &=&
      g_{1z} \left( \ba{cc} 0 & i   e^{ - i \phi_{\vec k}} W^{*2}\\
      - i   e^{  i \phi_{\vec k}}  W^2 & 0 \ea \right)
   \label{m1zeff}
   \eea
   The eigenvalues of this matrix are $\pm |W|^2$. We would
   like to find a new basis
   $ | \vec k, \alpha \rangle $ and $  | \vec k, \beta \rangle $
   such that the right hand side of \eref{m1zeff} has
   the same transformation of the $z$ component of a Pauli matrix.
   Evidently, we can just ``diagonalize" \eref{m1zeff} (so that $m_{1z}
   \propto \rho_z$)
   if we take
   $ | \vec k, \alpha \rangle \propto
    | \vec k, \alpha' \rangle
    - i e^{  i \phi_{\vec k}} \frac{W^2}{|W|^2} | \vec k, \beta' \rangle $.
   Keeping $ | \vec k, \beta \rangle = TP | \vec k, \alpha \rangle$,
   we have then
   \be
   m_{1z} (\vec k) = g_{1z} |W|^2 \rho_z
   \ee
   where we are using $\rho_{x,y,z}$ to denote Pauli matrices
   in $| \vec k, \alpha \rangle$ and $| \vec k, \beta \rangle$
   space.   Now $\rho_z$ has the correct transformation property.
   The overall phase factor for $| \vec k, \alpha \rangle$
   must further be chosen so that $\rho_{x,y}$ also transform correctly.
   The necessary choice can be seen to be
   \be
   | \vec k, \alpha \rangle = \frac{1}{\sqrt{2}} \left(
   \frac{W^*}{|W|}  | \vec k, \alpha' \rangle
    - i e^{  i \phi_{\vec k}} \frac{W}{|W|}| \vec k, \beta' \rangle
    \right)
    \label{ka}
    \ee
    and hence (via \eref{TP})
   \be
   | \vec k, \beta \rangle = \frac{1}{\sqrt{2}} \left(
    - i e^{  -i \phi_{\vec k}} \frac{W^*}{|W|}| \vec k, \alpha' \rangle
     +  \frac{W}{|W|}  | \vec k, \beta' \rangle
    \right)
    \label{kb}
    \ee
  With this choice, we have then
  \bea
  m_{1x} (\vec k) &=& g_{1p} |W|^2 \rho_x  \\
  m_{1y} (\vec k) &=& g_{1p} |W|^2 \rho_y
  \eea
    ensuring correct transformation properties of $\rho_{x,y,z}$.
  From \eref{ka}, we can verify explicitly
  \be
  P_+ | \vec k, \alpha \rangle =
  \frac{|W|}{\sqrt{2}}
    \left( \ba{c} 1 \\    1 \ea \right)
   \left( \ba{c} 1 \\  0 \ea \right)
   \ee
   hence involves only spin up $|s_z = 1\rangle$.  This basis
   thus satisfy the criteria of MCBB of \cite{Fu15}.
   One can indeed verify explicitly from \eref{ka} that we have
    (in the basis $|\sigma_z, s_z \rangle = $
   $|1,1\rangle \ $, $|1,-1\rangle\ $, $|-1,1\rangle\ $, $|-1,-1\rangle$)
   \be
   |\vec k, \alpha \rangle = \frac{1}{2 E_{\vec k} [1 + \frac{m}{E_{\vec k}}]^{1/2}}
   \left( \ba{c}
   E_{\vec k} + m - i v_z k_z \\
   i v k_\| e^{ i \phi_{\vec k}} \\
   E_{\vec k} + m + i v_z k_z \\
   - i v k_\| e^{ i \phi_{\vec k}} \ea \right)
   \ee
   which is
    then identical with that given in \cite{KF15,VKF16}.

   Since the magnetic moment is a crucial quantity determining
   many physical properties of the system, let us also obtain
   $\vec m_{2}$ in terms of $\vec \rho$.  We note
   that $m_{2z} \propto P_{-} s_z$, and similarly for the other
   components.  For later convenience, let us define ({\it c.f.} \eref{W})
      \be
   W_{2, \vec k} \equiv \frac{(E_{\vec k} + v k_\|) -
     (m + i v_z k_z)}{\sqrt{2} \mathcal{N}_{\vec k}}
     \label{W2}
     \ee
     We easily find
     \be
     |W_{2, \vec k}| = [\frac{1}{2} ( 1 - \frac{m}{E_{\vec k}})] ^{1/2}  \ .
     \ee
     Note that $|W_{\vec k}|^2 + |W_{2, \vec k}|^2 = 1$ for any $\vec k$.
     As before, we can replace $E_{\vec k}$ by the chemical potential
     $\mu$ for states on the Fermi surface, and shall again drop
     the subscripts $\vec k$ if no confusion can arise.
     At the Fermi level for electron doped Bi$_2$Se$_3$, we have
     thus $|W_2| > 1/\sqrt{2}$.

     We have, using \eref{ka1}, \eref{kb1} and \eref{ka},
    \be
   P_{-} | \vec k, \alpha \rangle = \frac{1}{2}
   \left( \ba{c} 1 \\ -1 \\ \ea \right)
    \left(
   \frac{W^* W_2}{|W|}  | \hat k + \rangle
    + i e^{  i \phi_{\vec k}} \frac{W W_2^*}{|W|}| \hat k - \rangle
    \right)
    \label{P2ka}
    \ee
    and
   \be
   P_{-} | \vec k, \beta \rangle = \frac{1}{2}
   \left( \ba{c} 1 \\ -1 \\ \ea \right)
    \left(
    - i e^{  -i \phi_{\vec k}} \frac{W^* W_2}{|W|}| \hat k + \rangle
     -  \frac{W W_2^*}{|W|}  |\hat k - \rangle
    \right)
    \label{P2kb}
    \ee
     where $|\hat k \pm \rangle$ are the spin wavefunctions defined
     in \eref{k+} and \eref{k-}.
     Defining the phase $\chi_{\vec k}$ by
     \be
     e^{ i \chi_{\vec k}} = \frac{ W W_2^*}{ | W| |W_2|}
     \label{chi}
     \ee
     We have $\chi_{\vec k} = \tan^{-1} \frac{ v_z k_z}{v k_\|}$, and is
    thus basically the polar angle of $\vec k$ if we rescale
    the anisotropic Fermi surface to a sphere.
     From \eref{P2ka} and \eref{P2kb} and noting
     $P_{-} s_{x,y,z} = P_{-} s_{x,y,z} P_{-}$, we easily find
     \be
     m_{2z} (\vec k) = g_{2z} |W_2|^2
     \left[ - \cos ( 2 \chi_{\vec k}) \rho_z
       + \sin ( 2 \chi_{\vec k})
       \left( \ba{cc} 0 & e^{- i \phi_{\vec k}} \\
       e^{ i \phi_{\vec k}} & 0 \ea \right) \right]
     \label{m2z}
     \ee
   \bea
     m_{2x} (\vec k) &=& g_{2p} |W_2|^2
     \left[ - \sin ( 2 \chi_{\vec k}) \cos (\phi_{\vec k}) \rho_z
       - \sin \phi_{\vec k}
       \left( \ba{cc} 0 &  - i e^{- i \phi_{\vec k}} \\
       i e^{ i \phi_{\vec k}} & 0 \ea \right)  \right. \nonumber \\
         & & \qquad \left.   - \cos ( 2 \chi_{\vec k}) \cos \phi_{\vec k}
       \left( \ba{cc} 0 &   e^{- i \phi_{\vec k}} \\
        e^{ i \phi_{\vec k}} & 0 \ea \right)\right]
     \label{m2x}
     \eea
        \bea
     m_{2y} (\vec k) &=& g_{2p} |W_2|^2
     \left[ - \sin ( 2 \chi_{\vec k}) \sin (\phi_{\vec k}) \rho_z
       + \cos \phi_{\vec k}
       \left( \ba{cc} 0 &  - i e^{- i \phi_{\vec k}} \\
       i e^{ i \phi_{\vec k}} & 0 \ea \right) \right. \nonumber \\
       & &   \qquad  \left.  - \cos ( 2 \chi_{\vec k}) \sin \phi_{\vec k}
       \left( \ba{cc} 0 &   e^{- i \phi_{\vec k}} \\
        e^{ i \phi_{\vec k}} & 0 \ea \right)\right]
     \label{m2y}
     \eea
     The above results are more transparent if we introduce
     \bea
     \rho_r &=& \cos \phi_{\vec k} \rho_x + \sin \phi_{\vec k} \rho_y \\
     \rho_{\phi} &=& -\sin \phi_{\vec k} \rho_x + \cos \phi_{\vec k} \rho_y
     \eea
     as the ``radial" and ``azimuthal" components of $\vec \rho$
     and similarly for $\vec m_{2}$.
     We have
     \be
     m_{2z} (\vec k) = g_{2z} |W_2|^2
     \left[ - \cos ( 2 \chi_{\vec k}) \rho_z + \sin ( 2 \chi_{\vec k}) \rho_r
     \right]
     \ee
      \be
     m_{2r} (\vec k) = g_{2p} |W_2|^2
     \left[ - \sin ( 2 \chi_{\vec k}) \rho_z - \cos ( 2 \chi_{\vec k}) \rho_r
     \right]
     \ee
     and
       \be
     m_{2 \phi} (\vec k) = g_{2p} |W_2|^2
     \left[  \rho_{\phi}     \right]
     \ee
    with the symmetries manifest  (note that $\chi_{\vec k}$ is odd
    under $k_z \to - k_z$).

    The form of $\vec m_{2}$ is somewhat more complicated than
    $\vec m_{1}$, which unfortunately is the price we have to
    pay since we used $\vec m_{1}$ to construct our basis.
    We note that, according to \cite{Liu10}, $g_{2z}$ and $g_{2p}$
    can in some instances be large compared with $g_{1z}$ and $g_{1p}$,
    for example for Sb$_2$Te$_3$.

    \subsection{another alternative basis}\label{subsec:m2}

    As mentioned, we can choose any quantity that transform as
    a pseudovector to aid our construction of the pseudospins.
    As an example, we here try using instead $\frac{1 - \sigma_x}{2}
    \vec s = P_{-} \vec s$.   Following the same procedure as
    subsection \ref{subsec:m1}, we first evaluate
    \be
      P_- |\vec k, \alpha' \rangle = \frac{W_2}{\sqrt{2}}
    \left( \ba{c} 1 \\
     -1 \ea \right)
    \left( \ba{c}
   1 \\
   i e^{i \phi_{\vec k}} \ea \right)
   \label{P2ka1}
   \ee
   and
         \be
    P_- |\vec k, \beta' \rangle = - \frac{W_2^*}{\sqrt{2}}
    \left( \ba{c} 1 \\
     -1 \ea \right)
    \left( \ba{c}
   i e^{-i \phi_{\vec k}} \\
    1 \ea \right)
    \label{P2kb1}
    \ee
    The matrix for $P_{-} s_z$ then becomes, in
    $|\vec k, \alpha' \rangle$ and $\vec k, \beta' \rangle$ space
    \bdm
          \left( \ba{cc} 0 &  - i   e^{ - i \phi_{\vec k}} W_2^{*2}\\
       i   e^{  i \phi_{\vec k}}  W_2^2 & 0 \ea \right)
      \edm
    With the same logic as before, we see that the proper choice
    for $|\vec k, \alpha \rangle_2$ (we distinguish our new basis
    in this section by the subscript $2$) should be
       \be
   | \vec k, \alpha \rangle_2 = \frac{i}{\sqrt{2}} \left(
   \frac{W_2^*}{|W_2|}  | \vec k, \alpha' \rangle
    + i e^{  i \phi_{\vec k}} \frac{W_2}{|W_2|}| \vec k, \beta' \rangle
    \right)
    \label{ka2}
    \ee
    and 
    from $| \vec k, \beta \rangle_2 \equiv
    TP | \vec k, \alpha \rangle_2$, we have
   \be
   | \vec k, \beta \rangle = - \frac{i}{\sqrt{2}} \left(
     i e^{  -i \phi_{\vec k}} \frac{W_2*}{|W_2|}| \vec k, \alpha' \rangle
     +  \frac{W_2}{|W_2|}  | \vec k, \beta' \rangle
    \right)
    \label{kb2}
    \ee
    Note the sign changes as well as the extra factors of
    $\pm i$ compared with \eref{ka} and \eref{kb}.
    Explicitly, we have
   \be
  P_- | \vec k, \alpha \rangle_2 = i
  \frac{|W_2|}{\sqrt{2}}
    \left( \ba{c} 1 \\    -1 \ea \right)
   \left( \ba{c} 1 \\  0 \ea \right)
   \ee
      \be
  P_- | \vec k, \beta \rangle_2 =  i
  \frac{|W_2|}{\sqrt{2}}
    \left( \ba{c} 1 \\    -1 \ea \right)
   \left( \ba{c} 0 \\  1 \ea \right)
   \ee
   which results in simple expressions for $\vec m_2$:
   \bea
   m_{2z} (\vec k) &=& g_{2z} |W_2|^2 \rho_{2z} \\
   m_{2x} (\vec k) &=& g_{2p} |W_2|^2 \rho_{2x} \\
   m_{2y} (\vec k) &=& g_{2p} |W_2|^2 \rho_{2y}
   \eea
   where we have denoted Pauli matrices in
   $| \vec k, \alpha \rangle_2|$$ \vec k, \beta \rangle_2$
   space by $\rho_{2,x,y,z}$.
   This basis has thus the advantage of having simpler
   forms for the magnetic moments if $g_{2z}$ and $g_{2p}$
   dominate over $g_{1z}$ and $g_{1p}$, or if $|W_2|$ is
   large compared with $|W|$.   However there is
   a disadvantage.  Explicit calculation shows that,
   for $\vec k$ residing in the northern hemisphere,
     \be
   |\vec k, \alpha \rangle_2 = \frac{i}{2 E_{\vec k} [1 - \frac{m}{E_{\vec k}}]^{1/2}}
   \left( \ba{c}
   E_{\vec k} - m + i v_z k_z \\
   i v k_\| e^{ i \phi_{\vec k}} \\
  - ( E_{\vec k} + m - i v_z k_z) \\
    i v k_\| e^{ i \phi_{\vec k}} \ea \right)
   \ee
   Since $|\vec k, \alpha \rangle_2$ in the southern hemisphere
   is dictated by $| - \vec k, \alpha \rangle_2 = P |\vec k, \alpha \rangle_2$,
   one can check that this pseudspin basis for the single
   particle has a discontinuity (sign change)
   across the equator, hence extra caution is necessary when
   calculations are made which involves, e.g., scattering between
   these states.  Note however the Cooper pair wavefunction,
   when expressed in terms of this basis, would {\em not} suffer
   from any artificial discontinuity across the equator,
   since the Cooper pairs involve both $\vec k$ and $-\vec k$.

\section{Conclusion}

In conclusion, we have presented a slight generalization
of the method in Ref \cite{Yip13} for constructing the pseudospin basis of
a time-reversal and inversion symmetric system.  We have
also shown how it can generate the ``manifestly covariant Bloch
basis" of Ref \cite{Fu15,KF15,VKF16} in the example
given by \eref{HN}.  We emphasize once more that the method of \cite{Yip13}
is applicable beyond \eref{HN}, in particular it is not restricted to the number
of orbitals involved.

\section{Acknowledgements}

This research is supported by the Ministry of Science and Education of
of Taiwan under grant numbers MOST-103-2119-M001-011-MY2 and
   MOST-104-2112-M-001-006-MY3.

\end{document}